\documentclass[a4paper]{jpconf}
\usepackage{graphicx}
\begin{document}
\title{10 Years of Object-Oriented Analysis on H1}

\author{Paul Laycock}

\address{Department of High Energy Physics, Oliver Lodge Laboratory,
University of Liverpool, Liverpool, L69 7ZE, UK}

\ead{laycock@hep.ph.liv.ac.uk}

\begin{abstract}
Over a decade ago, the H1 Collaboration decided to embrace the
object-oriented paradigm and completely redesign its data analysis
model and data storage format.  The event data model, based on the
RooT framework, consists of three layers - tracks and calorimeter
clusters, identified particles and finally event summary data - with a
singleton class providing unified access.  This original solution was
then augmented with a fourth layer containing user-defined objects.

This contribution will summarise the history of the solutions used,
from modifications to the original design, to the evolution of the
high-level end-user analysis object framework which is used by H1
today.  Several important issues are addressed - the portability of
expert knowledge to increase the efficiency of data analysis, the
flexibility of the framework to incorporate new analyses, the
performance and ease of use, and lessons learned for future projects.
\end{abstract}

\section{Introduction}

The H1 experiment analysed electron-proton data from the HERA machine
at DESY, Hamburg.  During the luminosity upgrade at the turn of the
millennium, H1 decided to change their analysis framework to an
object-oriented paradigm, concentrating on the data storage
model~\cite{H1OO1,H1OO2,H1OO3,H1OO4}.

\section{The Analysis Models}

\subsection{The old analysis model}

\begin{figure}
\begin{center}
\includegraphics[width=0.7\textwidth]{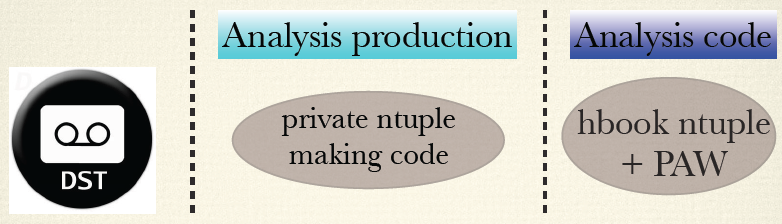}
\end{center}
\caption{\label{Fig:old}The old data analysis model.}
\end{figure}

Figure $\ref{Fig:old}$ shows the old data analysis model of H1.  The
Data Summary Tape (DST) fortran data storage format was processed
privately by private ntuple production code to produce the analysis
data format.  This was usually the hbook format, and physics analysis
was performed using PAW.  The scope for sharing analysis information
was fairly limited, as the variable definitions depended on the
privately maintained ntuple production code which varied from group to
group.  Meanwhile, the private production of ntuples was usually quite
inefficient, resulting in many copies of essentially the same data,
wasting resources.

\subsection{The H1OO data analysis model}

\begin{figure}
\begin{center}
\includegraphics[width=0.7\textwidth]{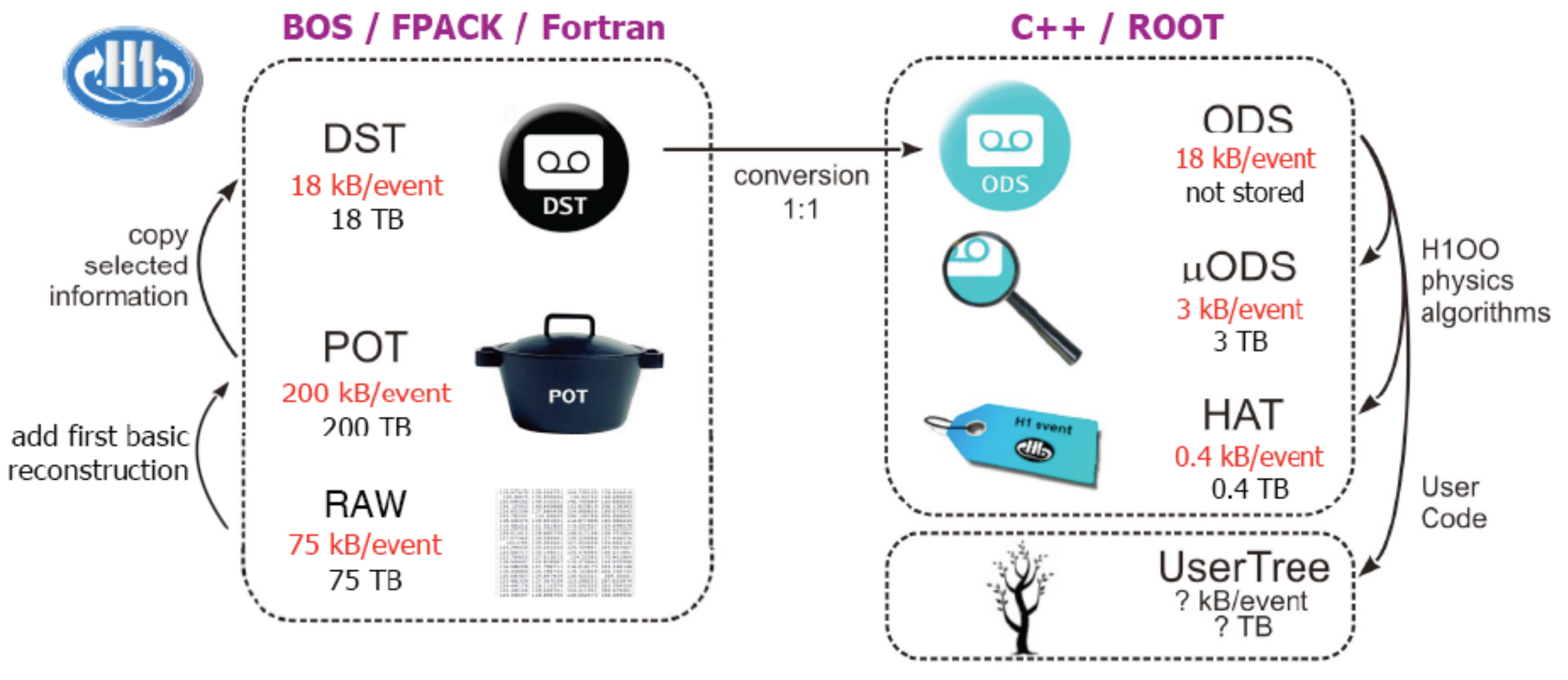}
\end{center}
\caption{\label{Fig:H1OO}The new data storage model of H1.}
\end{figure}

Figure $\ref{Fig:H1OO}$ shows the object-oriented data analysis model.
The aim was to centrally produce the analysis data storage format,
replacing the custom ntuples used previously.  Private ntuple
production was not prohibited and was used by some individuals, but
the vast majority of the collaboration enjoyed and still enjoys
central production of their analysis data format.

\subsection{The H1OO data storage model}

The DST remains the source of all derived analysis formats.  The
object-oriented data format consists of several layers, and all data
storage layers are encapsulated in the H1Tree interface.  This object
provides smart access to the data, such that the user doesn't need to
know which layer is accessed.

The Object Data Storage (ODS) layer is entirely equivalent to the DST,
but now Track and Cluster classes are used for the data
representation.  Object-oriented bank classes also exist, providing
DST, POT and RAW data access if and as necessary.  The ODS was
generally not used for analysis, as the content of the two derived
layers described next was usually found to be sufficient.  However, it
can be produced and stored if necessary (if the ODS content will be
repeatedly analysed), or created on-the-fly from DST (more performant
for rare access).

\begin{figure}
\begin{center}
\includegraphics[width=0.7\textwidth]{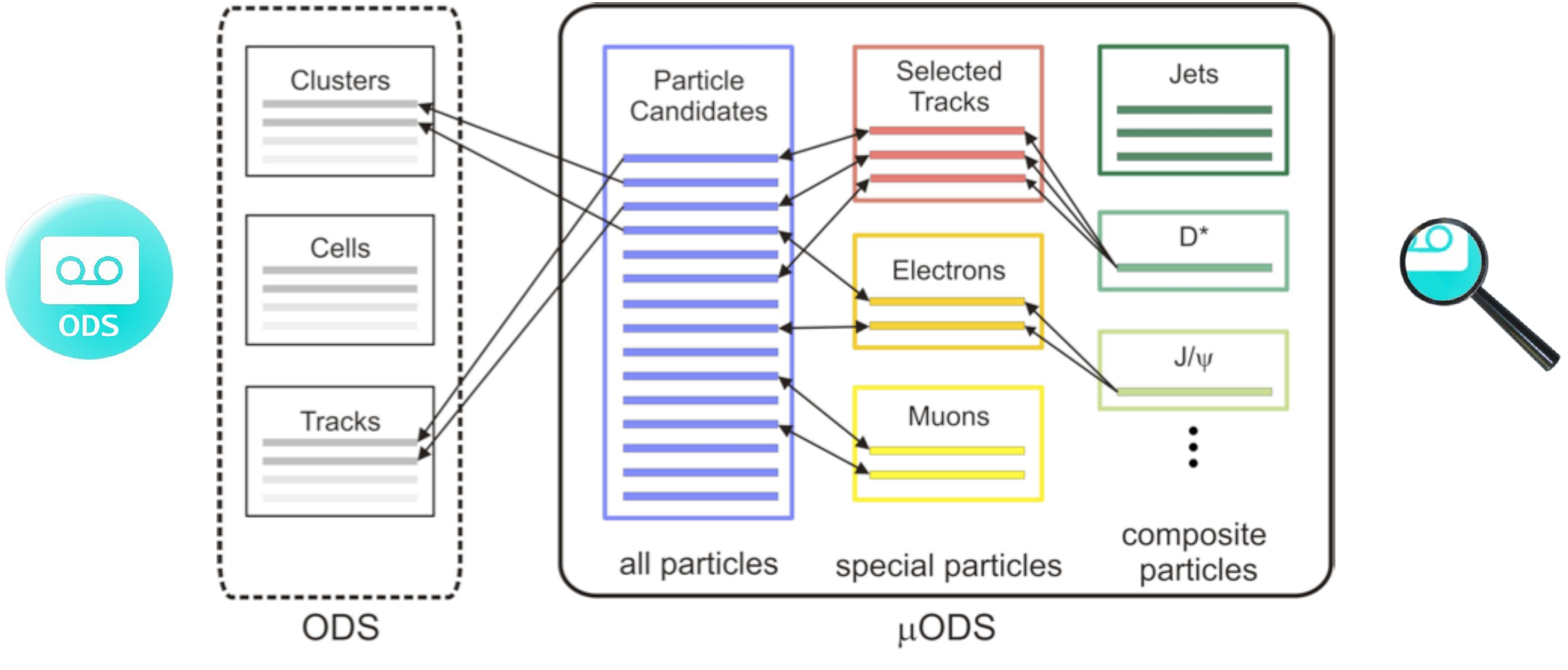}
\end{center}
\caption{\label{Fig:Part}The Particle Finder model.}
\end{figure}

Particle finders, shown in Figure $\ref{Fig:Part}$, are used to
produce the content stored in the next object-oriented layer, the
micro-ODS or MODS.  The particle-finding algorithms used are the best
knowledge of H1 by definition, and thus all analyses use this best
knowledge.  Event summary information is stored in the third layer,
the H1 Analysis Tag or HAT.  Together, the MODS and HAT layers are the
analysis data format, centrally produced, for the vast majority of H1
analyses.

A fourth and final optional layer, the UserTree, allows ultimate
flexibility by allowing a user-defined storage layer.  If this layer
was found to be useful for several groups, it entered the central
production framework and was also centrally maintained.  It's worth
noting that there are currently three UserTree packages in the core
H1OO framework, suggesting that this flexibility was both necessary
and useful.

\section{Transient Data}

The original H1OO concept concentrated on unifying the data analysis
format and optimising resource usage for it's production, and this was
very successful.  However, there were one or two limitations to this
model.  Chief among these was the lack of treatment of transient data,
i.e. quantities calculated or re-calculated at run time.  This was
found to be particularly relevant for the evaluation of systematic
uncertainties, where derived quantities need to be re-calculated based
on a systematically shifted base quantity.  It was also relevant when
analysts wanted to apply the latest calibrations on-the-fly, rather
than wait for a full-scale production of data and Monte Carlo.

\subsection{Transient data interface}

\begin{figure}
\begin{center}
\includegraphics[width=0.7\textwidth]{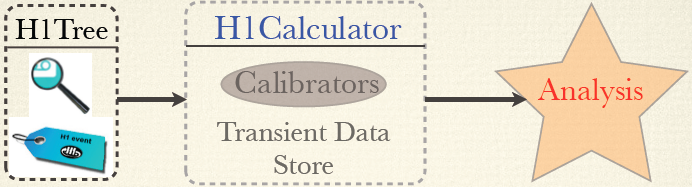}
\end{center}
\caption{\label{Fig:calc1}The transient data interface, H1Calculator.}
\end{figure}

To address these problems, a transient data interface called the
H1Calculator was designed and implemented~\cite{me, dave}, as shown in
Figure $\ref{Fig:calc1}$.  This reads the persistent data provided by
the H1Tree, and stores transient derived quantities.  The latest
calibration can then be applied easily, and systematic uncertainty
evaluation simply requires recalculating the derived quantities.  All
data access in user analysis code goes through this interface,
guaranteeing a consistent treatment of the data.

\begin{figure}
\begin{center}
\includegraphics[width=0.7\textwidth]{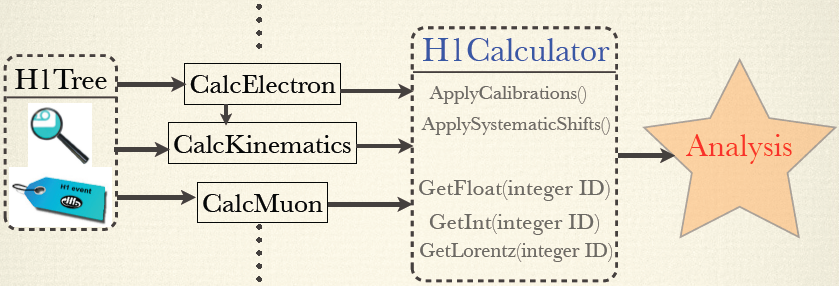}
\end{center}
\caption{\label{Fig:calc}A more detailed view of the transient data interface.}
\end{figure}

In practice, the scope for derived quantities for all H1 analyses is
very large, and the H1Calculator is composed of several smaller,
themed calculator classes which deal with specific quantities,
e.g. one for electron quantities, another for event kinematics.  A
generic interface to the data provides access to variables by type
(integer, TLorentzVector, etc.), which then allows user classes to be
decoupled from the details of this structure.  The main H1Calculator
class itself then provides a simple interface to variables by integer
ID (the generic interface) as well as switches to apply calibrations
and systematic shifts, as shown in Figure $\ref{Fig:calc}$.

\section{Higher Level Analysis Objects}

\begin{figure}
\begin{center}
\includegraphics[width=0.8\textwidth]{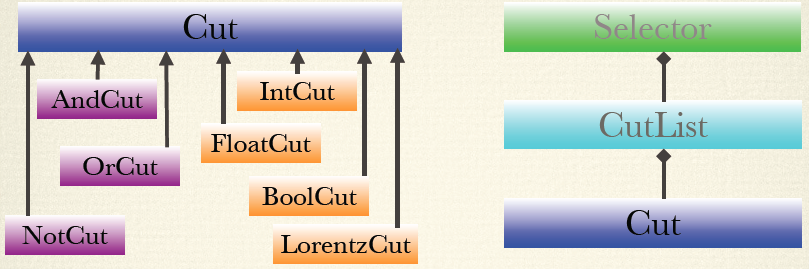}
\end{center}
\caption{\label{Fig:sel}The event selector classes.}
\end{figure}

Following the generic interface to the transient data, more end-user
classes could be defined.  One particular highlight are the event
selector classes~\cite{me}, composed of lists of cut objects.  A cut
object returns a boolean answer based either on a variable read from
the transient data interface or on a logical combination of other cut
objects.  This simple but very useful set of classes is shown in
Figure $\ref{Fig:sel}$.  They also provide detailed debugging
information and cutflow statistics.  These classes could be passed
from analyst to analyst to apply particular event selections, together
with other simple classes responsible for histogram
management~\cite{me}, i.e. classes which book and fill sets of
(related) histograms.  These simple organisational aides proved to be
very useful, not least in the context of data quality and software
validation, as well as in physics analysis.

\subsection{Analysis objects}

\begin{figure}
\begin{center}
\includegraphics[width=0.65\textwidth]{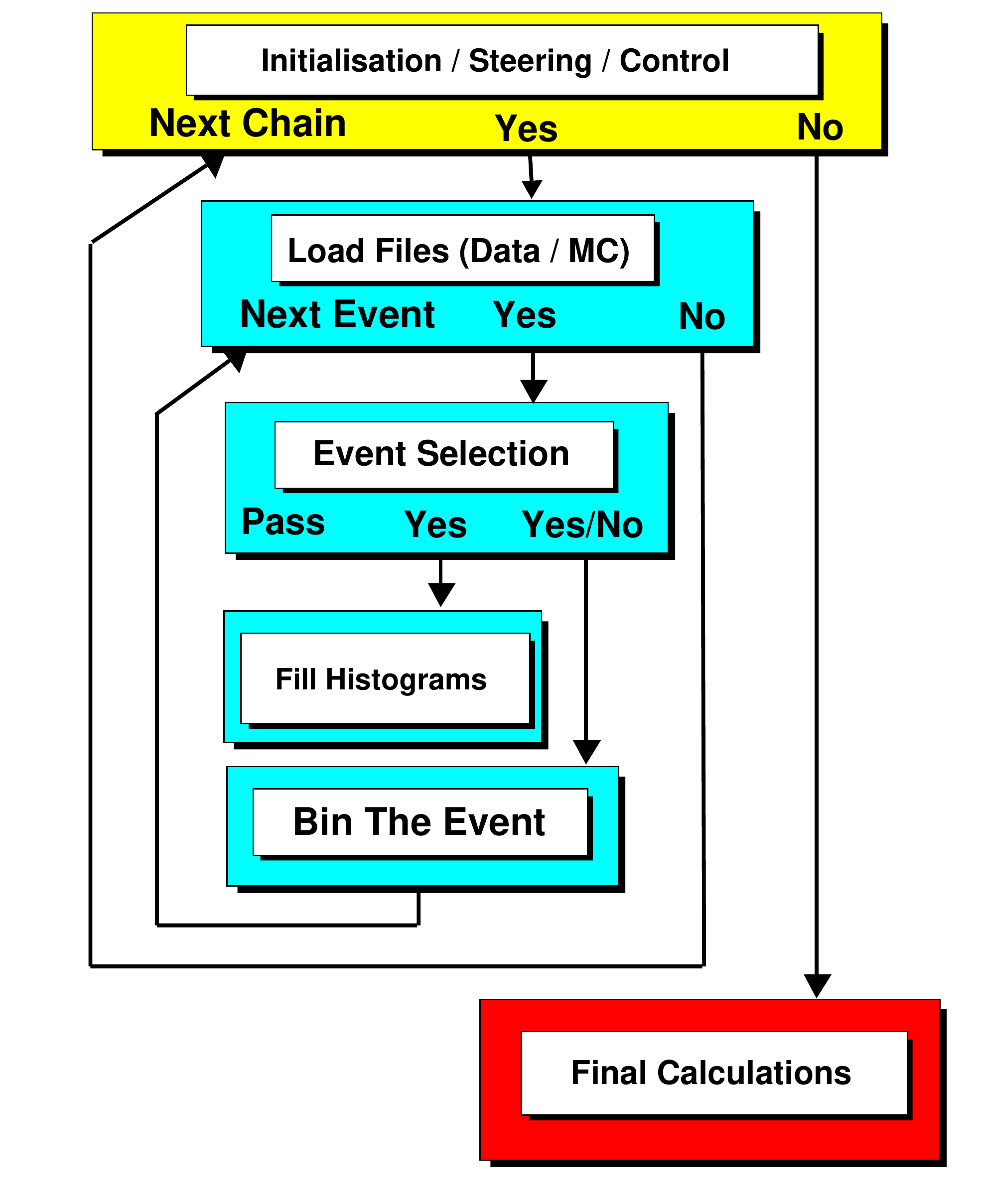}
\end{center}
\caption{\label{Fig:flow}Flow diagram of physics analysis.}
\end{figure}

Figure $\ref{Fig:flow}$ shows a flow diagram of a simplified physics
analysis.  Identifying the H1Calculator as ``the event'', the selector
and histogram manager objects are also evident.  Two other
organisational structures can also be seen, namely at the level of
looping over one set of data or Monte Carlo, where we used the term
``chain'' from PAW, and finally the analysis level itself.

\begin{figure}
\begin{center}
\includegraphics[width=0.6\textwidth]{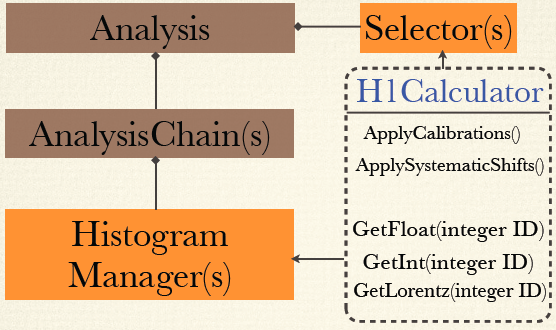}
\end{center}
\caption{\label{Fig:analysis}Analysis object model.}
\end{figure}

Correspondingly, Analysis and AnalysisChain objects also proved to be
useful organisational classes~\cite{me}.  An Analysis object is
composed of several AnalysisChains, which each contain one or more
Histogram Managers.  The Event Selector is an object defined at the
Analysis level.  Data access goes through the H1Calculator.  This
simple model is shown in Figure $\ref{Fig:analysis}$ and allows a
common framework for nearly all stages of a physics analysis.  This in
turn allowed better collaboration between physics groups, and the
easy exchange of high-level analysis code.

\section{Conclusions}

During the luminosity upgrade at the turn of the millennium, H1
decided to change their analysis framework to an object-oriented
paradigm, concentrating on the data storage model.  This move was very
successful in unifying the data storage model and analysis formats of
H1.  The flexibility to have a user defined data storage layer proved
crucial in several analyses.

The development of a transient data interface improved physics
analysis in many cases, especially those where access to the latest
calibrations was critical and/or complicated systematic effects had to
be evaluated.  It paved the way for further developments which allowed
for a more efficient exchange of higher-level physics analysis tools,
up to and including entire analyses.

\ack The material presented here is the culmination of many years of
work by members of the H1 Collaboration.  I'd like to thank all of
those who worked on the H1OO project and in particular my friend and
collaborator Dave South.

%\subsection{Acknowledgments}
%command \verb"\ack" sets the acknowledgments heading as an unnumbered
%section.

\end{document}